\documentstyle[12pt]{article}
\begin{document}
\begin{titlepage}

\title{Semiclassicality and Decoherence of Cosmological Perturbations}
\author{David Polarski$^{\dag \ddag}$ and Alexei A. Starobinsky$^{\S}$\\
\hfill \\
\dag~Laboratoire de Mod\`eles de Physique Math\'ematique, EP93 CNRS\\
Universit\'e de Tours, Parc de Grandmont, F-37200 Tours (France)\\
\hfill\\
\ddag~D\'epartement d'Astrophysique Relativiste et de Cosmologie,\\
Observatoire de Paris-Meudon, 92195 Meudon cedex (France)\\
\hfill \\
\S~Landau Institute for Theoretical Physics, \\
Kosygina St. 2, Moscow 117334 (Russia) \\}

\date{20 April 1995}
\maketitle

\begin{abstract}
Transition to the semiclassical behaviour and the decoherence process for
inhomogeneous perturbations generated from the vacuum state during an
inflationary stage in the early Universe are considered both in the Heisenberg
and the Schr\"odinger representations to show explicitly that both approaches
lead to the same prediction: the equivalence of these quantum
perturbations to classical perturbations having stochastic Gaussian amplitudes
and belonging to the quasi-isotropic mode. This equivalence and the
decoherence are achieved once the exponentially small (in terms of the
squeezing parameter $r_k$) decaying mode is neglected. In the
quasi-classical limit $|r_k|\to \infty$, the perturbation mode functions can
be made real by a time-independent phase rotation,
this is shown to be equivalent to a fixed relation between squeezing angle
and phase for all modes in the squeezed-state formalism. Though the present
state of the gravitational wave background is not a squeezed quantum state
in the rigid sense and the squeezing parameters loose their direct meaning
due to interaction with the environment and other processes, the
standard predictions for the rms values of the perturbations generated during
inflation are not affected by these mechanisms
(at least, for scales of interest in cosmological applications). This
stochastic background still occupies a small part of phase space.
\end{abstract}

\end{titlepage}

\section{Introduction}

A unique and remarkable property of the inflationary scenario of the early
Universe (irrespective of its concrete realization) is that it opens an
exciting possibility to directly observe the outcome of a genuine quantum-
gravitational effect: generation of quasi-classical fluctuations of quantum
fields including the gravitational one in strong external gravitational
fields (in other words, by space-time curvature). Historically, this effect
was known as "particle creation from the vacuum in a background gravitational
field", but it is clear now that what can be measured at present are not
"particles" but rather inhomogeneous fluctuations (perturbations) of the
gravitational field. In non-inflationary cosmological models, the effect of
particle creation is exceedingly small and does not lead to observable
consequences. Just the opposite, not only can minimal perturbations of the
gravitational potential generated in the simplest versions of the inflationary
scenario (first quantitatively calculated in~\cite{hsg}) be sufficient to
explain galaxy formation and the large-scale structure in the Universe, but
also their predicted spectrum (approximately flat, $n\simeq 1$) and statistics
(Gaussian) have been successfully quantitatively confirmed by the $COBE$
discovery of large-angle fluctuations $\frac{\Delta T}{T}$ of the cosmic
microwave background temperature~\cite{smoot} 10 years after the prediction
was made. Moreover, in the case of the inflationary scenario and in contrast
with other cases, the corresponding "pure" quantum-gravitational effect~-~
creation of gravitons by background gravitational fields~-~produces a large
relic gravitational-wave background with frequencies $\ll 10^{10}{\rm Hz}$ in
the Universe~\cite{staro79}, and it is even possible that a modest, but still
significant part of the observed large-angle $\frac{\Delta T}{T}$ fluctuations
(at the level proposed in~\cite{staro85}, but probably not larger) is due to
these gravitational waves. Another possibility of observing creation of
particles by gravitational fields might be through the Hawking radiation from
primordial black holes (PBH) with masses $M\leq 10^{15}{\rm g}$, but it
follows from
direct or indirect observational tests (see, e.g.~\cite{carr}) that the number
density of such PBH in the Universe is very small if they formed at all (and
the inflationary scenario typically predicts their complete absence).

In spite of this definite success of the inflationary scenario and of the
quantum
theory in curved space-time, some confusion still seems to exist in the
literature regarding how rigid the derivation of the perturbations is (see,
e.g. polemics in~\cite{grish1,grish2,albrecht}. The key problem here is that
though the process of creation from the vacuum and the perturbations
themselves are purely of a quantum-mechanical nature (at least initially), the
observed temperature or density fluctuations in the Universe are certainly
classical.

Thus a complete derivation should include some mechanism of
quantum-to-classical transition and decoherence of the perturbations.
Connected with this are fundamental questions about the wave function of
the Universe being pure
or mixed and its interpretation. An additional complication is the relation
between the Heisenberg and the Schr\"odinger representations in quantum
mechanics and quantum field theory. Of course, these representations are
completely equivalent and contain the same physics, but they use a different
language and different parameters for the description of a given state of a
quantum field. Almost all initial studies of particle creation in cosmology
in general~\cite{par69,zel,grish3}, and in the inflationary scenario in
particular~\cite{hsg,staro79} were performed using the Heisenberg
approach. This approach is more convenient for the purpose of renormalization
{}~\cite{zel,par74} and a description of the quantum-to-classical
transition (we shall come back to the latter point below). The use of mode
functions satisfying a classical wave equation and the Bogolubov
transformation for creation and annihilation operators is a characteristic
feature of this approach. Derivation of the perturbations in this approach
usually ends up (like in~\cite{hsg,staro79}) by taking these mode
functions as classical variables with stochastic Gaussian amplitudes but
satisfying a certain condition in the regime outside the Hubble radius
("non-decreasing modes"). On the other hand what naturally follows from quantum
cosmology where a homogeneous isotropic background is quantized too, is the
Schr\"odinger representation for the wave functions of perturbations (see
e.g.~\cite{hal}). This representation is usually used also in order to
consider the decoherence process. Here one speaks about a two-mode squeezed
state and describes it with the help of squeezing parameters. Though, of
course, it is generally well-known that the Bogolubov transformation of
annihilation and creation operators in the Heisenberg picture just corresponds
to the evolution of the vacuum state into a squeezed one in the Schr\"odinger
picture (see e.g. a
mathematical analysis in~\cite{barut,perelomov,schum}), there still exists
a point
of discussion about how the two approaches are related in the cosmological
context and which of them is "better". Inspired by the impression that a
squeezed state has a non-classical behaviour even for large values of the
squeezing parameter $r$, there were even claims that the squeezed state
formalism gives observable predictions which are superior to the usual
Heisenberg approach~\cite{grish1,grish2} (actually it does not, see
also~\cite{albrecht}). It is clear that a deep understanding of the generation
process of perturbations is of utmost importance both for further development
of the theory of quantum gravity and quantum cosmology and for observational
implications. That is why we reconsider this question here.
\par
We will deal with the simplest case of a quantum real massless scalar field
$\phi$ in a Friedman-Robertson-Walker (FRW) background because it is
sufficiently representative for our purposes. First, the equation for the
time-dependent part of $\phi$ coincides exactly with the equation for the
time-dependent part of gravitational waves on a FRW background~\cite{lifshits}
under the condition that the non-diagonal components of the matter pressure
tensor are zero in the first order of perturbation expansion. The latter
condition is satisfied e.g. in the case of matter consisting of a mixture of
an arbitrary number of hydrodynamical components with barotropic equations of
state $p_i=p_i(\epsilon_i)$ and scalar fields with arbitrary mutual
interactions and minimal coupling to the gravitational field. Furthermore, this
equation has a structure very similar (though generally not completely
identical) to the equation satisfied by the gravitational potential.

In section 2, we remind the reader of explicit relations between mode
functions of the field, coefficients of the Bogolubov transformation and
squeezing parameters. Then, in section 3, we consider the quasi-classical
limit in the Heisenberg and Schr\"odinger representations in parallel in order
to emphasize the fact that a two-mode squeezed state for large absolute
values of the squeezing parameter $|r|$ is completely equivalent to a
classical {\em standing} wave with a {\em stochastic} Gaussian amplitude. The
term "standing" means that there is a definite deterministic correlation
between ${\bf k}$ and $-{\bf k}$ modes for each wave vector ${\bf k}$. Also,
we show how this property is related to the fundamental fact that the field
modes can be made real by a time-independent phase rotation in this limit.
All this is illustrated with a specific but very important example, namely
that of the de Sitter background. In section 4, we remind the physical process
in the Universe
leading to extreme squeezing ($|r|\to \infty$) and to the quantum-to-classical
transition - the different behaviour of non-decreasing and decaying modes
outside the Hubble radius. Then it follows that quantum-to-classical
transition and decoherence in the Heisenberg representation (in contrast with
the Schr\"odinger one) are achieved simply by omitting an exceedingly small
part of the field operator (the decaying mode), without any need to consider
some interaction of the mode with an "environment". This may be called,
following J. A. Wheeler's favourite way to put it, {\em "decoherence without
decoherence"}. After this omission, it becomes unimportant whether the
field is in a
pure or in a mixed state. As a result, we come to the conclusion that the
Heisenberg (field mode) approach becomes more straightforward in real
situations when
very small interactions of the field with other fields take place. Namely, due
to these interactions and the resulting decoherence process, the present
quantum state of the field $\phi$ is neither a pure squeezed state nor even
can it be described by a squeezed density matrix, so the use of squeezing
parameters looses sense. On the other hand, interactions practically do not
change the field modes (at least for sufficiently large scales). Thus, all
predictions about present-day perturbations remain unchanged. A possibility
of having nevertheless some "quantum signature" in the present-day spectrum of
perturbations is mentioned. Finally, we discuss the point that the omitted,
exponentially small part of the field may be important for the calculation of
the entropy of the perturbations.

\section{Bogolubov transformation and two-mode squeezed state}

We give here the essential about quantized fields on a flat FRW background.
Let us consider a real massless scalar field $\phi$.
It is described by the following action $S$:
\begin{equation}
S={1\over 2}\int d^4x \sqrt{-g}\partial^{\mu}\phi \partial_{\mu}\phi\label{S}
\end{equation}
where $\mu=0,..,3,~c=\hbar=1$ and the Landau-Lifshitz sign conventions are
used. The space-time metric has the form
\begin{equation}
ds^2=dt^2-a^2(t)\delta_{ij}dx^i dx^j, \qquad\ i,j=1,2,3.
\end{equation}
Let us remind how the dynamics of this system will lead to the appearance
of squeezed states.  We first write down the classical Hamiltonian
$H$ in terms of the field $y\equiv a\phi$ and the conformal time $\eta =
\int {dt \over a(t)}$. The following result is then obtained
\begin{eqnarray}
H &=& \int d^3{\bf x}~{\cal H}(y, p, \partial_i y, t)\nonumber\\
&=& {1\over 2}\int d^3{\bf k} \lbrack p({\bf k})p^*({\bf k})+k^2
y({\bf k})y^*({\bf k})+{{a'}\over a} \left(y({\bf k})p^*({\bf k})
+p({\bf k})y^*({\bf k})\right) \rbrack
\label{Hcl}
\end{eqnarray}
where
\begin{equation}
p\equiv{{\partial {\cal L}(y,y')}\over {\partial y'}}=y'-\frac{a'}{a}y
\end{equation}
and a prime stands for derivation with respect to the conformal time.
Here the following Fourier transform convention is used:
$\Phi({\bf k})\equiv (2\pi)^{-3/2}\int \Phi({\bf r})e^{-i{\bf kr}}
d^{3}{\bf r}$ for functions as well as for operators.
In order to avoid too heavy notations, we will often write simply $y({\bf k}),
a({\bf k}),...$ instead of $y({\bf k},\eta),a({\bf k},\eta),...$ though the
Fourier transforms are time-dependent c-functions or time-dependent
operators in the Heisenberg representation.
Due to reality of the field $y$, we have that $y({\bf k})=y^{*}
(-{\bf k})$, resp. $y^{\dag}(-{\bf k})$ for operators. Therefore,
any classical field configuration is completely specified by
giving the Fourier transforms in half Fourier space.
This may be not true in the quantum case, and the full Fourier space
has to be used if a quantum state of the field is not invariant
under the reflection ${\bf k}\to -{\bf k}$.  However, for the vacuum
initial state that we will use below, there is no such complication.
The Fourier transforms appearing in~(\ref{Hcl}) satisfy the equation
\begin{equation}
y''({\bf k})+\left(k^2-{a''\over a}
\right)y({\bf k})=0.  \label{eq}
\end{equation}

When the field $y$ is quantized, the Hamiltonian becomes
\begin{eqnarray}
H = \int {{d^3{\bf k}}\over 2}\bigl\lbrack k \bigl (a({\bf k})a^{\dag}
({\bf k})+a^{\dag}({\bf k})a({\bf k}) \bigr )+i{a'\over a}\bigl (a^{\dag}
({\bf k})a^{\dag}(-{\bf k})-
a({\bf k})a(-{\bf k}) \bigr )\bigr\rbrack .   \label{H}
\end{eqnarray}
The time-dependent (in the Heisenberg representation) operator $a({\bf k})$
appearing in~(\ref{H}) is defined as usual:
\begin{equation} a({\bf k})= {1\over
\sqrt{2}}\left(\sqrt{k}~y({\bf k}) + i {1\over \sqrt{k}} p({\bf k})\right),
\end{equation}
so that
\begin{equation} y({\bf k})={a({\bf k})+a^{\dag}(-{\bf
k})\over \sqrt{2k}},~~~
p({\bf k})=-i\sqrt{{k\over 2}}\left(a({\bf k})-a^{\dag}({-\bf k})\right).
\end{equation}
The canonical commutation relations
\begin{equation}
[y({\bf x},\eta)~,~p({\bf x}',\eta)]=i\delta^{(3)}({\bf x}-{\bf x}')\label{cc}
\end{equation}
imply the following commutation relations \begin{equation}
[y({\bf k},\eta)~,~p^{\dag}({\bf k}',\eta)]=i\delta^{(3)}({\bf k}-{\bf k}'),
\hspace{2truecm}
[a({\bf k},\eta)~,~a^{\dag}({\bf k}',\eta)]=\delta^{(3)}({\bf k}-{\bf k}')
\label{com}.
\end{equation}
The last piece in the integrand of~(\ref{H}) is responsible for the squeezing.
Let us see first how it affects the time evolution of the system. We have
\begin{equation}
\left(
\begin{array}{c}
a'({\bf k})\\
(a^{\dag}(-{\bf k}))'\\
\end{array}
\right)=
\left(
\begin{array}{cc}
-ik         &{a'\over a}\\
{a'\over a} &ik\\
\end{array}
\right)
\left(
\begin{array}{c}
a({\bf k})\\
a^{\dag}(-{\bf k})\\
\end{array}
\right).
\end{equation}
Clearly, the general solution of these two coupled equations are
\begin{eqnarray}
a({\bf k},\eta) &=& u_k(\eta) a({\bf k},\eta_0)+
v_k(\eta) a^{\dag}(-{\bf k},\eta_0),\nonumber\\
a^{\dag}(-{\bf k},\eta) &=& u_k^*(\eta) a^{\dag}(-{\bf k},\eta_0)+
v_k^*(\eta) a({\bf k},\eta_0)\label{a}.
\end{eqnarray}
This is just a Bogolubov transformation.
Eq.(\ref{a}) can be interpreted as giving the time evolution of the
creation and annihilation operators in the Heisenberg representation, or as a
definition of explicitly time-dependent operators in the Schr\"odinger
representation.
The commutation relations~(\ref{com}) are preserved under the unitary time
evolution which yields the constraint
\begin{equation}
|u_k(\eta)|^2-|v_k(\eta)|^2=1,\label{uv1}
\end{equation}
therefore, allowing the following standard parameterization of the
functions $u_k,v_k$:
\begin{eqnarray}
u_k(\eta)&=&e^{-i\theta_k(\eta)}\cosh r_k(\eta), \nonumber\\
v_k(\eta)&=&e^{i(\theta_k(\eta)+2\varphi_k(\eta))}\sinh r_k(\eta).\label{uv}
\end{eqnarray}
Here $r_k$ is the squeezing parameter, $\varphi_k$ is the squeezing
angle and $\theta_k$ is the phase.

The relation between these quantities and those introduced in the
$\alpha -\beta$ formalism~\cite{zel} is the following (if $\Omega_k$
in~\cite{zel} is chosen to be equal to $k$):
\begin{equation}
u_k=\alpha_ke^{-ik\eta},~~~v_k^*=\beta_ke^{ik\eta}.
\end{equation}
Also, the parameters $s_k,{\tilde u}_k,\tau_k$ introduced in \cite{zel}, that
proved to be very useful to make the adiabatic expansion for large $k$ and to
obtain a finite average value of the energy-momentum tensor of the quantum
field $\phi$ either by the $n$-wave regularization method \cite{zel}
or by the equivalent adiabatic regularization method \cite{par74}, are
expressed through the squeezing parameters as:
\begin{eqnarray}
s_k &=& |\beta_k|^2=\sinh^2r_k~, \nonumber \\
{\tilde u}_k &=& \alpha_k\beta_k^*e^{-2ik\eta}+\alpha^*_k\beta_ke^{2ik\eta}=
\cos 2\varphi_k \sinh 2r_k~, \nonumber \\
\tau_k &=& i(\alpha_k\beta_k^*e^{-2ik\eta}-\alpha_k^*\beta_ke^{2ik\eta})=
-\sin 2\varphi_k \sinh 2r_k \label{suv}
\end{eqnarray}
and we use here the notation ${\tilde u_k}$ to avoid confusion with the
quantity $u_k$ used in the present paper.
Note that $s_k$ is equal to the average number of created particles with
momentum ${\bf k}$
in the WKB regime (in particular, if $a(\eta)$ becomes constant).
We don't intend here to introduce the notion of particles in a non-WKB regime
because it is ambiguous and does not lead to interesting results.
\par
Let us now define the field modes $f_k(\eta)$ with $\Re f_k\equiv
f_{k1}$ and $\Im f_k\equiv f_{k2}$, $f_k(\eta_0)=1/\sqrt{2k}$,
we will adopt a similar notation for the quantities $p$ and $y$,
\begin{eqnarray}
y(\bf k)&\equiv& f_k(\eta) a({\bf k},\eta_0)+f_k^*(\eta)
a^{\dag}(-{\bf k},\eta_0)\nonumber\\
&=&\sqrt{2k}f_{k1}(\eta) y({\bf k},\eta_0)-\sqrt{{2\over k}}f_{k2}(\eta)
p({\bf k},\eta_0)\label{yk}
\end{eqnarray}
and the momentum modes $g_k(\eta)$,~ $g_k(\eta_0)=\sqrt{{k\over 2}}$,
\begin{eqnarray}
p(\bf k)&\equiv& -i\bigl\lbrack g_k(\eta) a({\bf k},\eta_0)
-g_k^*(\eta) a^{\dag}(-{\bf k},\eta_0)\bigr\rbrack\nonumber\\
&=&\sqrt{{2\over k}}g_{k1}(\eta) p({\bf k},\eta_0)+
\sqrt{2k}~g_{k2}(\eta)y({\bf k},\eta_0).\label{pk}
\end{eqnarray}
The modes $f_k$ satisfy the Euler-Lagrange equation
(\ref{eq}). Note that
\begin{eqnarray}
f_k &=& {u_k+v_k^*\over \sqrt{2k}},~~~~~|f_k|^2={1\over 2k}(\cosh 2r_k+
\cos 2\varphi_k \sinh 2r_k),\nonumber\\
g_k &=& \sqrt{\frac{k}{2}}(u_k-v_k^*)=i(f'_k-{a'\over a}f_k).\label{fg}
\end{eqnarray}
Eqs~(\ref{uv},\ref{fg}) give explicitly the relation between the modes $f_k$
which are typically used in the Heisenberg approach and the squeezing
parameters characteristic for the Schr\"odinger approach. Also, they can be
used to obtain the dynamical equations satisfied by the squeezing parameters,
see Eq.~(\ref{dyn}) below.
The Wronskian condition for Eq.(\ref{eq}), as well as the commutation
relations (\ref{com}), yield the following equality
\begin{equation}
g_kf_k^*+g_k^*f_k=i(f'_kf_k^*-f^{'*}_kf_k)=1. \label{wr}
\end{equation}
We will be interested, in particular, in the quantum state of the
field $y$ defined to be vacuum
at some time $\eta_0$ in the following way
\begin{equation}
\forall {\bf k}:~~ a({\bf k},\eta_0)|0,\eta_0\rangle = 0.
\end{equation}
This state corresponds to a Gaussian state and time evolution preserves
its Gaussianity.
Indeed it follows from~(\ref{yk},\ref{pk}) that in the Heisenberg
representation, the time independent state $|0,\eta_0\rangle_H$ is an
eigenstate of the operator $y({\bf k})+i\gamma_k^{-1}(\eta)p({\bf k})$, namely
\begin{equation}
\Bigl \lbrace y({\bf k})+i\gamma_k^{-1}(\eta)p({\bf k})
\Bigr \rbrace|0,\eta_0\rangle_H=0~,\label{eq1}
\end{equation}
where the operators $y({\bf k}),p({\bf k})$ as well as the function
$\gamma_k$ depend on time,
\begin{eqnarray}
\gamma_k &=& k~{u^*_k-v_k\over u^*_k+v_k}=k~\frac{1-i\sin 2\varphi_k
\sinh 2r_k}{\cosh 2r_k+\cos2\varphi_k \sinh2r_k}
=\frac{1}{2|f_k|^2}-i\frac{F(k)}{|f_k|^2}~,\nonumber\\
F(k) &=& \Im u_k v_k=\Im f_k^* g_k~=\frac{1}{2}\sin 2\varphi_k \sinh 2r_k~.
\label{Fk}
\end{eqnarray}
On the other hand, in the Schr\"odinger representation the time-evolved state
$|0,\eta\rangle_S\equiv S|0,\eta_0\rangle$, where $S$ is the $S$-matrix,
satisfies the equation
\begin{equation}
S a({\bf k},\eta_0) S^{-1}|0,\eta\rangle_S=0
\end{equation}
or equivalently
\begin{equation}
\Bigl \lbrace y({\bf k},\eta_0)+i\gamma_k^{-1}(\eta)p({\bf k},\eta_0)
\Bigr \rbrace|0,\eta\rangle_S=0~.\label{eq2}
\end{equation}
Note the similar structure of Eqs (\ref{eq1},\ref{eq2}).
In the coordinate Schr\"odinger representation, $p({\bf k},\eta_0)=
-i \partial/ \partial y(-{\bf k},\eta_0)$.
Hence the state $|0,\eta_0\rangle_S$ has a Gaussian wave functional
in this representation consisting of the product of
\begin{eqnarray}
\Psi[y({\bf k},\eta_0),y(-{\bf k},\eta_0)] &=& N_k\exp \left(
-{{y({\bf k},\eta_0)y(-{\bf k},\eta_0)}\over {2|f_k|^2}}\lbrace 1-i2F(k)
\rbrace \right) \nonumber \\
&=& N_k\exp \left(-\frac{|y({\bf k},\eta_0)|^2} {2|f_k|^2}
\lbrace 1-i2F(k)\rbrace \right)\label{Psi}
\end{eqnarray}
for each pair ${\bf k},-{\bf k}$
where $N_k$ is a normalization coefficient. The time dependence of $\Psi$ is
through $f_k,~F(k),~{\rm and}~N_k$.
This structure of the wave
functional just reflects the fact that we get a two-mode squeezed state.
The corresponding probability ${\cal P}[y({\bf k},\eta_0),y(-{\bf k},\eta_0)]$
is given by
\begin{equation}
{\cal P}[y({\bf k},\eta_0),y(-{\bf k},\eta_0)]\propto \exp \left(
-\frac{|y({\bf k},\eta_0)|^2}{|f_k|^2}\right)~.\label{pdf}
\end{equation}
At $\eta=\eta_0$, we have $\gamma_k(\eta_0)=k$ or equivalently $F(k)=0$,
in other words we have a minimum uncertainty wave function.

\section{Transition to semiclassical behaviour}

Let us first consider the transition in the Heisenberg approach and take
the formal limit "$\hbar \to 0$" keeping the rms amplitude $|f_k|$, when
expressed in physical units, fixed. Since the right-hand sides of the
commutation
relation~(\ref{cc}) and of Eqs~(\ref{uv1},\ref{wr}) are proportional to $\hbar$
in physical units and do not depend on $|f_k|$, they may be approximately
replaced by 0 in this limit. In other words, $|u_k|\approx |v_k|\gg 1,~|f_k|\gg
1/\sqrt{2k}, {}~|g_k|\gg \sqrt{\frac{k}{2}}$ in natural units in the quasi-
classical limit. Then it
follows from Eq.(\ref{wr}) with 0 in the right-hand side that $f^*_k=c_k f_k$,
where $c_k$ is a time-independent constant. As a result, it is possible to make
$f_k$ real for all times by a {\em time-independent} phase rotation, viz.
$f_k\to f_k \exp(-\frac{i}{2} {\rm arg}c_k)$. On the other hand, $g_k$ is
purely imaginary in this limit.
\par
A further consequence is that all variables $y({\bf k})$ and
$p({\bf k})$ become mutually commuting.  However, we still cannot ascribe any
definite numerical values to them, in contrast with coherent states in the
quasi-classical limit; there is no Bose condensate. The correct way to put it
is that field modes become equivalent to {\em stochastic} c-number functions of
time with some probability distribution $\rho (y({\bf k})y(-{\bf k}))
\equiv \rho (|y({\bf k})|^2)$ for each pair of modes ${\bf k},-{\bf k}$
in the following sense
\begin{equation}
_{H}\langle 0,\eta_0|G(y({\bf k}))G^{\dag}(y({\bf k}))|0,\eta_0\rangle_H =
\int \int dy_1({\bf k}) dy_2({\bf k}) \rho (|y({\bf k})|) |G(y({\bf k}))|^2
\end{equation}
for any arbitrary function $G\left(y({\bf k})\right)$ and ${\bf k}\not= 0$, 
(for ${k=0}$ the proof is a little bit different). Here we assume for 
simplicity 
of notation the ${\bf k}$ spectrum to be discrete, this can be achieved e.g. 
by formally making the $T^3$ identification of space with three topological 
comoving scales much larger than all scales of interest. By considering
average values of arbitrary powers of $y({\bf k})$ and $y^{\dag}({\bf k})$
and using the Wick theorem, it is straightforward to show
that the two-dimensional probability distribution $\rho (|y({\bf k})|)$ is
Gaussian with the dispersion $\langle y_1^2\rangle + \langle y_2^2 \rangle =
|f_k|^2$. Indeed, if
\begin{equation}
G(y)=\sum_{m=0}^{\infty} q_my^m~, ~~~~\rho
(|y|)={1\over \pi |f_k|^2} \exp \left(-{|y|^2\over |f_k|^2}\right)
\end{equation}
where the argument ${\bf k}$ of $y$ is omitted for brevity and $f_k$ has been
made real, then
\begin{eqnarray}
_{H}\langle0,\eta_0|G(y)G^{\dag}(y)|0,\eta_0\rangle_H =
\sum_{m=0}^{\infty}\sum_{n=0}^{\infty}q_mq_n^{\ast}f_k^{m+n}
{_H}\langle 0,\eta_0|\left(a({\bf k},\eta_0) + a^{\dag}(-{\bf k},\eta_0)
\right)^m   \nonumber \\
\left(a^{\dag}({\bf k},\eta_0)+a(-{\bf k},\eta_0)\right)^n
|0,\eta_0\rangle_H =  \sum_{m=0}^{\infty} m!~|q_m|^2f_k^{2m} = \nonumber \\
\sum_{m=0}^{\infty}\sum_{n=0}^{\infty}q_mq_n^{\ast}
\int \int dy_1dy_2 \rho (|y|)y^my^{\ast n} =
\int \int dy_1dy_2 \rho (|y|) |G(y)|^2.
\end{eqnarray}
Thus, in the continuous limit, the equivalent classical stochastic field 
can be written as $y({\bf k},\eta)=f_k(\eta)e({\bf k})$ where the quantities 
$e({\bf k})$ are
time-independent $\delta$-correlated Gaussian variables with zero average and
unit dispersion:  $\langle e({\bf k})e^*({\bf k}')\rangle=\delta^{(3)} ({\bf
k}-{\bf k}'),~e^*({\bf k})=e(-{\bf k})$.  The crucial property here is that the
time-dependent part $f_k(\eta)$ factorizes both in the operator part $a({\bf
k},\eta_0)+a^{\dag} (-{\bf k},\eta_0)$ of the field mode $y({\bf k},\eta)$ and
in the stochastic variable $e({\bf k})$ of the equivalent c-number function in
the limit involved. As a result, after some realization of the stochastic
amplitude of the field mode has occured, further evolution of the mode is
deterministic and is not affected by quantum noise.  This kind of
quantum-to-classical transition is similar to that used in the stochastic
approach to inflation~\cite{star84,star86}.  
\par 
We turn now to a description
of the same transition in the Schr\"odinger representation.
We see from~(\ref{Psi}) that semiclassicality is implied if the following
condition is satisfied \begin{equation} |F(k)|\gg 1\label{F}.
\end{equation} It is clear from~(\ref{F}) that this requires the quantum state
to be extremely squeezed, namely $|r_k|\gg 1$. Note that, in this limit, we
cannot omit the number 1 appearing inside the figure brackets of~(\ref{Psi})
because it would make the wave function non-normalizable.
The classicality is to be understood in the following sense: if we assign to
each point in phase-space $\bigl (y({\bf k}),~p({\bf k})\simeq \frac{F(k)}
{|f_k|^2}y({\bf k})\bigr )$
the probability given by~(\ref{pdf}), it will move with time according to the
classical Hamiltonian equations.
Writing $y({\bf k})=|y({\bf k})| e^{i\vartheta}$, it follows
from~(\ref{pdf}) that $|y({\bf k})|$ obeys a Rayleigh distribution
while the phase $\vartheta$ becomes a stochastic variable uniformly distributed
($0\leq \vartheta \leq 2\pi$). Condition~(\ref{F}) can also be expressed in
the following way
\begin{equation}
\frac{|\gamma_{k2}|}{|\gamma_{k1}|}\gg 1~,\hspace{2cm}\gamma_k\equiv
\gamma_{k1}+i\gamma_{k2}~.\label{gamma1}
\end{equation}
This can also be seen from the following vacuum expectation values
\begin{eqnarray}
\langle \Delta y({\bf k},\eta)\Delta y^{\dag}({\bf k}',\eta)\rangle &\equiv&
\Delta y^2(k)\delta^{(3)}({\bf k}-{\bf k}')=
|f_k|^2 \delta^{(3)}({\bf k}-{\bf k}')~,\nonumber\\
\langle \Delta p({\bf k},\eta)\Delta p^{\dag}({\bf k}',\eta)\rangle &\equiv&
\Delta p^2(k)\delta^{(3)}({\bf k}-{\bf k}')=
|g_k|^2 \delta^{(3)}({\bf k}-{\bf k}')~.
\end{eqnarray}
where $\Delta\Phi\equiv \Phi-\langle \Phi\rangle$ and we have further adopted
the notation $\langle \Phi({\bf k},\eta)
\Phi^{\dag}({\bf k}',\eta)\rangle \equiv
\Phi^2(k)\delta^{(3)}({\bf k}-{\bf k}')$, where the quantity $\Phi^2(k)$, the
power spectrum of the quantity $\Phi$ depends only on $k$ if the state is
invariant under spatial translations and rotations. Note further that
$\langle y\rangle = \langle p\rangle=0$.
The following identities can then be shown to hold, in complete accordance
with the wave functional representation~(\ref{Psi})~:
\begin{eqnarray}
\Delta y^2(k)\Delta p^2(k)&=&|f_k|^2 |g_k|^2\nonumber\\
&=&(f_{k1}g_{k2}-f_{k2}g_{k1})^2+{1\over 4}\label{fg1}\nonumber\\
&=&{1\over 4}(\sin^22\varphi_k \sinh^22r_k +1)=F^2(k)+{1\over 4}~.
\label{fg2}
\end{eqnarray}
Then the condition of semiclassicality~(\ref{F}) corresponds to an uncertainty
which is much bigger than the minimal one allowed by the rules of quantum
mechanics. This shows once more that we are dealing in this limit with
stochasticity of a classical type.
It is interesting to note that we also have the following vacuum expectation
values
\begin{eqnarray}
\langle \Delta y_1({\bf k},\eta)\Delta y_1({\bf k}',\eta)\rangle &=&
\frac{|f_k|^2}{2}\Bigl (\delta^{(3)}({\bf k}-{\bf k}')+\delta^{(3)}({\bf k}+
{\bf k}')\Bigr )~,\label{yp1}\\
\langle \Delta y_2({\bf k},\eta)\Delta y_2({\bf k}',\eta)\rangle &=&
\frac{|f_k|^2}{2}\Bigl (\delta^{(3)}({\bf k}-{\bf k}')-\delta^{(3)}({\bf k}+
{\bf k}')\Bigr )~,\\
\langle \Delta p_1({\bf k},\eta)\Delta p_1({\bf k}',\eta)\rangle &=&
\frac{|g_k|^2}{2}\Bigl (\delta^{(3)}({\bf k}-{\bf k}')+\delta^{(3)}({\bf k}+
{\bf k}')\Bigr )~,\\
\langle \Delta p_2({\bf k},\eta)\Delta p_2({\bf k}',\eta)\rangle &=&
\frac{|g_k|^2}{2}\Bigl (\delta^{(3)}({\bf k}-{\bf k}')-\delta^{(3)}({\bf k}+
{\bf k}')\Bigr )~.\label{yp2}
\end{eqnarray}
The equalities~(\ref{yp1}-\ref{yp2}) express the correlation existing between
${\bf k}$ and $-{\bf k}$ modes. They are in complete agreement with the fact
that for any real quantity $f$ with
$f({\bf k})\equiv f_1({\bf k})+if_2({\bf k})$, we have $f_1({\bf k})=
f_1(-{\bf k}),~f_2({\bf k})=-f_2(-{\bf k})$.
\par
Let us prove now that $f_k$ can be made real for $|r_k|\to \infty$ using the
squeezed state formalism. When we write the equation of motions for
$\theta_k$ and $\varphi_k$, the following equations are obtained
\begin{eqnarray}
r_k' &=& \frac{a'}{a} \cos 2\varphi_k~,\nonumber\\
\varphi_k' &=& -k-\frac{a'}{a}\coth 2r_k \sin 2\varphi_k~,\nonumber\\
\theta_k' &=& k+\frac{a'}{a}\tanh r_k \sin 2\varphi_k~.\label{dyn}
\end{eqnarray}
The first two equations are coupled and their solutions can then be substituted
in the last equation. However, it is interesting to combine the last two
equations, yielding the following property
\begin{equation}
\lim_{|r_k| \to \infty} (\theta_k+\varphi_k)'= 0~.
\end{equation}
This means that $\theta_k + \varphi_k \to \delta_k$, where $\delta_k$ is some
constant phase. This in turn implies that the field modes $f_k$ have the
following asymptotic behaviour for $r_k\to \infty$
\begin{equation}
\sqrt{2k}f_k\to e^{-i\delta_k}e^{r_k}\cos \varphi_k~,
\end{equation}
and analogously for the modes $g_k$
\begin{equation}
\sqrt{{2\over k}}g_k\to ie^{-i\delta_k}e^{r_k}\sin
\varphi_k~.
\end{equation}
The quantity $\delta_k$ can be made zero by a time-independent phase rotation
in which case $f_k$ becomes real. It is interesting to note that~(\ref{F})
implies
\begin{equation}
\lim_{|r_k| \to \infty} \sin 2\varphi_k \sinh 2r_k=\infty
\end{equation}
even if $\sin 2\varphi_k\to 0$.  This is also in accordance with the fact
$\langle n_k\rangle=\sinh^2r_k\delta^{(3)}({\bf k}-{\bf k}')$ so that large
$r_k$ alone is enough in order to have semi-classicality.
\par
One more way to study the quantum-to-classical transition is through
the Wigner function formalism. The Wigner function for the two-mode
squeezed state ${\bf k},-{\bf k}$ is defined as
\begin{equation}
W\left(y({\bf k}),y(-{\bf k}),p({\bf k}),p(-{\bf k})\right)=
{1\over (2\pi)^2}\int \int dx_1dx_2\, e^{-i(p_1x_1+p_2x_2)}
\langle y({\bf k})-{x({\bf k})\over 2}|~\hat \rho~|y({\bf k})
+ {x({\bf k})\over 2}\rangle
\end{equation}
where $\hat \rho$ is the density matrix of the state
and $y_1$ and $y_2$ are the real and imaginary parts of $y({\bf k})$
in the Schr\"odinger coordinate represenation, and the same convention applies
to $x({\bf k})$ and $p({\bf k})$.
Using
$\hat \rho = |\Psi \rangle \langle \Psi|$ with $\Psi$ given by (\ref{Psi}),
we obtain
\begin{eqnarray}
W({\bf k},-{\bf k})={1\over (2\pi)^2}\int \int dx_1dx_2\, e^{-i(p_1x_1+p_2x_2)}
\Psi^{\ast}\left(y({\bf k})-{x({\bf k})\over 2}\right)
\Psi \left(y({\bf k})+{x({\bf k})\over 2}\right) = \nonumber \\
N_k^2{|f_k|^2\over \pi}\exp \left(-{|y({\bf k})|^2\over |f_k|^2}\right)
\exp \left(-|f_k|^2|p({\bf k})-{F(k)\over |f_k|^2}y({\bf k})|^2\right).
\label{wig}
\end{eqnarray}
$W>0$ in this case, so there is no problem with its interpretation as a
probability distribution in phase space. As explained above, the
quantum-to-classical transition is achieved by taking the formal limit
$\hbar \to 0$ keeping the physical amplitude $|f_k|$ fixed. Then
the Wigner function becomes
\begin{eqnarray}
W({\bf k},-{\bf k})=N_k^2\exp \left( -{|y({\bf k})|^2\over |f_k|^2}\right)
\delta \left (p_1({\bf k})-{F(k)\over |f_k|^2}y_1({\bf k})\right)
\delta \left (p_2({\bf k})-{F(k)\over |f_k|^2}y_2({\bf k})\right) = \nonumber \\
{\cal P}[y({\bf k}),y(-{\bf k})]\, \delta \left(p({\bf k})-{F(k)\over |f_k|^2}
y({\bf k})\right).  \label{redwig}
\end{eqnarray}
This just describes the deterministic motion in phase space of a bunch of
trajectories with stochastic Gaussian initial amplitude .
\par
The Wigner function (\ref{redwig}) coincides with the probability
distribution in phase space that was obtained in~\cite{GP85} for a toy
model of an upside-down harmonic oscillator. We see that this behaviour
of the Wigner function is general in the large-squeezing limit and does not
rely neither on assumptions made in~\cite{GP85}, nor on the existence of the
inflationary stage. In the paper~\cite{LM93} an objection was raised
against this way of making the quantum-to-classical transition based on
the observation that it is always possible to find new canonical variables
$\tilde y ({\bf k}),\tilde p ({\bf k})$ for which $\langle \tilde y^2
\rangle \langle \tilde p^2\rangle ={1\over 4}$ (even in the limit
$|r_k|\to \infty$). However, it is clear from the previous discussion
that this objection is not relevant because the requirement
$\langle \tilde y^2\rangle \langle \tilde p^2\rangle \gg 1$ for {\em all}
possible canonically conjugate variables is not the necessary condition
for the semi-classical behaviour. The only necessary condition is
$|r_k|\gg 1$, or Eq.(\ref{52}) below for the physical amplitude of
gravitational waves.
\par
In inflationary theories, primordial perturbations are generated by
vacuum quantum fluctuations of a real scalar field where the power spectrum of
the quantum field fluctuations is given by $|f_k|^2$.
Let us consider the very important example of a massless real scalar field on
a (quasi) de Sitter space. In that case we have
\begin{equation}
\sqrt{2k}f_k=e^{-ik\eta}\bigl (1-\frac{i}{k\eta}\bigr )~,\hspace{1cm}
\sqrt{\frac{2}{k}}g_k=e^{-ik\eta}~,\hspace{1cm}
\eta\equiv -\frac{1}{aH}<0~.\label{dS}
\end{equation}
The modes~(\ref{dS}) give also a very accurate description for slowly varying
Hubble parameter $H$, namely when $|{\dot H}|\ll 3H^2$.
After some straightforward calculation we get
\begin{equation}
u_k=e^{-i(k\eta+\delta_k)}\cosh r_k \hspace{1cm}v_k=e^{i(k\eta+\frac{\pi}{2})}
\sinh r_k\label{u}
\end{equation}
and $\sinh r_k=\frac{1}{2k\eta}\to -\infty~{\rm when}~k\eta \to 0$.
The crucial point in~(\ref{u}) is that $\tan\delta_k=\frac{1}{2k\eta}$,
therefore $\delta_k$ will tend to the constant value $-\frac{\pi}{2}$. Hence
in the limit $k\eta\to 0$, the modes $f_k$ are purely real up to a
constant phase transformation. Let us give for completeness the solution for
$\varphi_k$ and $\theta_k$
\begin{eqnarray}
\varphi_k &=& \frac{\pi}{4}-\frac{1}{2}\arctan \frac{1}{2k\eta}~,\nonumber\\
\theta_k &=& k\eta+\arctan \frac{1}{2k\eta}~.
\end{eqnarray}
We have finally
\begin{equation}
F(k)=\frac{1}{2}\sin 2\varphi_k \sinh 2r_k \sim (2k\eta)^{-1}\to -\infty
\end{equation}
for $k\eta \to 0$ though $\sin 2\varphi_k \to 0$.

\section{Long-wave mode behaviour and decoherence}

The physical mechanism producing the Bogolubov transformation and the extreme
squeezing in the Universe is, as well known, the expansion of the Universe
(not necessarily inflation) and
the existence of the Hubble radius $R_H\equiv H^{-1}=\frac{a^2}{a'}$. For
modes with $kR_H\ll a$, i.e. with wavelengths outside the Hubble radius, the
general solution of Eq.~(\ref{eq}) has the following form in terms of the mode
functions $f_k$, with $y({\bf k})$ expressed through $f_k$ using~(\ref{yk})
{}~\cite{staro79}:
\begin{equation}
f_k=C_1({\bf k})a+C_2({\bf k}) a \int_{\infty}^{\eta} \frac{d\eta'}
{a^2(\eta')}~,\hspace{1cm}g_k={\cal O}(iC_1({\bf k})k^2a\eta)~ +
i\frac{C_2({\bf k})}{a}~. \label{LW}
\end{equation}
$C_1({\bf k})$ can be made real by a phase rotation. Then the normalization
condition leads to the relation
\begin{equation}
C_1 \Im C_2=-\frac{1}{2}~.\label{CC}
\end{equation}
The first term
in~(\ref{LW}) is the quasi-isotropic mode, also called the growing mode. It
corresponds to a constant value of the field $\phi$. The same behaviour is
shared by the leading term of scalar (adiabatic) metric perturbations in the
synchronous gauge for an arbitrary scale factor $a(\eta)$ as well as by the
gravitational potential $\Phi$ during stages of power-law expansion.
The name quasi-isotropic means that this mode does not spoil the isotropic
expansion of the Universe at early stages, furthermore it is contained in
the linear expansion of the Lifshits-Khalatnikov quasi-isotropic
solution~\cite{LL}.
The second term in~(\ref{LW}) is the decaying mode.
\par
In inflationary models, comparing~(\ref{dS}) with~(\ref{LW}), we get
\begin{equation}
C_1=\frac{H_k}{\sqrt{2k^3}}~,\hspace{1cm}C_2=-\frac{ik^{3/2}}{\sqrt{2}H_k}
\label{C1C2}
\end{equation}
in agreement with~(\ref{CC}) where $H_k$ is the value of the slowly varying
Hubble parameter $H$ at the moment of the first Hubble radius crossing
$\eta_1~(\eta_1<0, |\eta_1(k)|=k^{-1})$. Here we have multiplied $f_k$
in~(\ref{dS}) by $-i$ to make it both real and positive.
Then both terms in~(\ref{LW}) are of the same order at
$\eta\sim \eta_1$. After that, the decaying mode quickly becomes exceedingly
small. For example, for scales of interest for cosmological applications that
crossed the Hubble horizon about $60-70$ e-folds before the end of inflation,
the rms of the decaying mode is $e^{-2r_k}\sim 10^{-80}$ or less from that of
the quasi-isotropic mode at the end of inflation, and it becomes $<10^{-95}$
up to the present moment. It is clear that we should neglect the decaying
mode completely. One more formal reason for this is that one cannot keep
small terms of the relative order of $e^{-2r_k}$ as far as renormalization
contributions to each mode, which are of the order of $e^{-r_k}$, are not
taken into account.
\par
But once the second term in the expression for $f_k$ in~(\ref{LW}) is omitted,
we obtain immediately decoherence because quantum coherence is described by
the correlation~(\ref{CC}) between the non-decaying {\em and} decaying modes.
Therefore, when working with the field modes $f_k(\eta)$ in the Heisenberg
representation, there is no need to consider any interaction with an
"environment" and trace over its degrees of freedom in order to get
decoherence. Moreover, after neglecting the decaying mode, it becomes
unimportant whether the quantum state of a given mode ${\bf k}$ (or the
Universe as a whole) is pure of mixed since the difference between the three
following field configurations, viz. classical stochastic field with modes
$f_k$ given by~(\ref{LW}), pure squeezed quantum state satisfying~(\ref{LW},
\ref{CC}) and mixed squeezed quantum state with the relation~(\ref{CC})
understood in the sense of rms values for $C_1(\bf k)$ and $C_2(\bf k)$, is
exponentially small ($\sim e^{-2r_k}$) and disappears after this omission.
Summarizing, in the {\em peculiar} case of quantum cosmological perturbations
generated during inflation, the decoherence can be obtained without
consideration of any concrete decoherence process, that is why we may call it
{\it "decoherence without decoherence"}.
\par
This property makes the Heisenberg approach more straightforward and convenient
in real physical situations when one takes into account small interactions
of the perturbations with matter in the Universe. For example, it is known
that the interaction of scalar perturbations and gravitational waves with
a background matter having shear viscosity does not change the quasi-isotropic
mode but yields an additional exponential decay of the decaying mode. On the
other hand, new perturbations belonging to both modes may be generated by
local physical processes, especially after the second Hubble radius crossing
during the radiation- or matter-dominated FRW stages. Their amplitude is much
more than the fantastically small amplitude of the decaying mode remaining
from the inflationary stage but much less
(at least for scales exceeding $R_{eq}\simeq 15 h^{-2}{\rm Mpc}$
where $h$ is the present value of the Hubble constant in terms of 100km/s/Mpc)
than the amplitude of the quasi-isotropic mode. As a
result, the present state of the perturbations is (of course) neither a pure
squeezed state, nor does it make sense to call the corresponding density
matrix "squeezed" because one has no reasons to expect to find any direction
in phase space where the noise (uncertainty) would be less or even comparable
to the quantum limit $\frac{\hbar}{2}$ for the vacuum state. A much more
adequate description of this state is that it consists of the classical
stochastic part described by the quasi-isotropic modes $f_k$ (the first term
in Eq.(\ref{LW}) for $k|\eta|\ll 1$ or its continuation to the regime $k\eta
\geq 1$, see below) plus some small noise of indefinite structure.
Then the squeezing parameters loose their original sense. One may still
formally introduce them through the mode functions using Eqs~(\ref{uv},
\ref{fg}) as their definitions (as is done, e.g. in Eq.(\ref{50}) below),
but they will have little if any relation to the squeezed quantum state in a
narrow, rigid sense.
\par
Note that our method of omitting the decaying mode or, equivalently, taking
the "$\hbar \to 0$" limit with $|f_k|$ being fixed in physical units
may be also thought of as coarse graining, or smoothing, in phase
space. Namely, the radius of the coarse graining should be chosen in such
a way as to smooth completely the "subfluctuant" variable $p({\bf k})-
{F_k\over |f_k|^2}y({\bf k})$ (using the terminology of~\cite{schum})
but still without affecting the "superfluctuant" variable $y({\bf k})$
significantly. After that, this radius may be put zero in all problems
concerning dynamical evolution and stochastic properties of perturbations
(except for the calculation of their entropy). However, it is clear from
the previous discussion that final results do not depend on any concrete
type of coarse graining as far as the leading quasi-isotropic mode
remains unaffected by it.
\par
Let us now consider what happens to the quasi-isotropic mode after the second
Hubble radius crossing. Then one has $f_k=C_1({\bf k})f_q(k,\eta)$ where
$f_q$ is the exact solution of Eq.(\ref{eq}) with the asymptotic behaviour
$f_q(k,\eta)=a(\eta)$ for $k\eta\ll 1$, in particular,
$f_q(k,\eta)=\frac{a_0}{k} \sin k\eta$
during the radiation-dominated regime $a=a_0 \eta$. Note that
$f_q=A(k)\sin (k\eta+\xi_k)$ for $k\eta\gg 1$ and an arbitrary behaviour
of $a(\eta)$. Here $\xi_k$ is a constant and $A(k)\sim a(\eta_2)$ where
$\eta_2$ is the moment of second Hubble radius crossing, $\eta_2\sim k^{-1}$.
It is this constant phase
$\xi_k$, or the phase $\theta_k$ linearly growing with conformal time
$\eta$ (see~(\ref{51}))
that one usually has in mind when saying that the gravitational waves
generated during inflation have stochastic amplitudes but fixed phases. The
fact that $y({\bf k})$ becomes zero at some moments of time shows that these
waves are standing ones. The adiabatic perturbations do not have such an
oscillating
behaviour at the matter-dominated stage. Hence, it is more general to speak
not about the fixed phase of perturbations but about the type of the
mode (the quasi-isotropic one).
\par
This demonstrates that the loss of quantum coherence
does not preclude the existence of strong quasi-classical correlations, c.f.
comparison of quantum coherence and classical correlation in~\cite{morikawa}
(see also~\cite{habib} in this connection). If we formally introduce the
parameters $r_k, \theta_k, \varphi_k$ using~(\ref{uv},\ref{fg}), then
\begin{eqnarray}
e^{2r_k} &=& 2 k C_1^2 \Bigl\lbrack f_q^2+\frac{a^2}{k^2}((f_q/a)')^2
\Bigr\rbrack~,\hspace{1cm}r_k\gg 1~,\nonumber\\
\varphi_k &=& -\theta_k~,~~~\sin\theta_k=-\frac{\frac{a}{k}(\frac{f_q}{a})'}
{\sqrt{f_q^2+\frac{a^2}{k^2}\bigl ((\frac{f_q}{a})'\bigr )^2}}\label{50}
\end{eqnarray}
for the quasi-isotropic mode. Note the important persisting correlation between
$\varphi_k$ and $\theta_k$. Deep inside the Hubble radius, $\eta\gg
\eta_2(k)$,
\begin{equation}
e^{2r_k}\simeq 2 k C_1^2 A^2 = {\rm constant}~,\hspace{1cm}\theta_k=
k\eta+\xi_k-\frac{\pi}{2}~.\label{51}
\end{equation}
It is in this regime that it becomes possible to introduce the number of
created particles (gravitons) $n({\bf k})=n(-{\bf k})\simeq \frac{e^{2r_k}}
{4}$.
The condition for the semiclassical behaviour of gravitational waves
inside the Hubble radius expressed directly in terms of their amplitude
$h_{ij}=-\frac{\delta g_{ij}}{a^2}\sim \sqrt{G}\phi$ looks like
\begin{equation}
k^3h_g^2(k)\equiv \langle k^3 h_{ij}({\bf k})h^{ij}({\bf k})\rangle
\sim e^{2r_k}\frac{l_P^2}{\lambda^2}\gg \frac{l_P^2}{\lambda^2}\label{52}
\end{equation}
where $l_P$ is the Planck length and $\lambda=2\pi a k^{-1}$ is the
wavelength of the perturbations. The corresponding condition for adiabatic
perturbations is the same at the radiation-dominated stage inside the Hubble
radius, and it is even less restrictive at the matter-dominated stage. Thus, if
primordial perturbations are measurable at all, they are always classical;
their quantum origin is reflected in their power spectrum and statistics
only. It is important to emphasize here that if primordial perturbations are
quasi-isotropic at present, this does not {\em necessarily} imply their
quantum origin from a squeezed state. It is just the opposite: {\em any}
classical or quantum process that does not spoil the isotropy of the Universe
at sufficiently early times results in the dominance of the quasi-isotropic
mode nowaday. The role of the inflationary scenario here is to provide a
causal mechanism for the generation of the perturbations, while in a non-
inflationary cosmology assuming FRW behaviour of the early Universe, power
spectrum and statistics of the quasi-isotropic mode may be chosen
arbitrarily by hand. On the contrary, if the present GW background was
generated at late times, after the second Hubble radius crossing
(e.g. by cosmic strings), then both modes would be present with equal
probability and the phase $\xi_k$ would be uniformly distributed. This,
as well known, gives a possibility to discriminate between primordial
and non-primordial GW.

Is it possible to verify experimentally the predicted quasi-isotropic
character of the primordial gravitational wave background on sufficiently
large scales, i.e. the standing wave behaviour of $f_k$, or that of
$\theta_k$ in~(\ref{51})? A direct LIGO-type experiment is clearly
hopeless because it
would require a frequency resolution at the level of the Hubble constant, i.e.
$10^{-18}$Hz. It is remarkable, however, that this prediction is already
proved by observations in the following sense.
Let us assume that the present gravitational-wave background is of primordial
origin and its spectral energy density is at least
\begin{eqnarray}
\frac{\lambda}{\epsilon_c}\frac{d\epsilon_g(\lambda)}{d\lambda}& \sim &
10^{-10} \Bigl (\frac{\lambda}{R_H} \Bigr )^2~,\hspace{1cm}R_{eq}<\lambda
<R_H~,\nonumber\\
& \sim & 10^{-14}\hspace{2.5cm}\lambda_{\gamma}\ll \lambda<R_{eq}\label{GWB}
\end{eqnarray}
($\epsilon_c=3H_0^2/8\pi G$ is the critical energy density, $R_H=H_0^{-1}=
3000 h^{-1}{\rm Mpc},~\lambda_{\gamma}=2\pi T_{\gamma}^{-1}
\sim 1{\rm cm}$), as expected in the simplest versions of the
inflationary scenario, or larger (the latter is possible for
$\lambda < 100 h^{-1}{\rm Mpc}$). This background corresponds to the value
$k^{3/2}h_g(k)\sim 10^{-5}$ at the second Hubble radius crossing. Then
it follows from the CMB $\frac{\Delta T}{T}$ data that this background should
be dominated by the quasi-isotropic modes $\propto f_q(k)$ for scales
$\lambda>100h^{-1}{\rm Mpc}$ (corresponding to a multipole number $l<40$).
Furthermore, the success of
the primordial nucleosynthesis theory proves convincingly that the Universe
was isotropic beginning from $t\sim 1$s ($T_{\gamma}\sim 1{\rm MeV}$). As a
consequence, the present background~(\ref{GWB}) should
be dominated by the quasi-isotropic modes for $\lambda >h^2 R_{eq}\sim 15
{\rm Mpc}~(t(\eta_2)\sim 10^{10}{\rm s})$.
This region can be further expanded using observational limits on the PBH
number density~\cite{carr}. Of course, the
inflationary scenario predicts the dominance of the quasi-isotropic mode for
all scales $\gg 1{\rm cm}$ for the primordial background. As explained above,
this quasi-isotropic behaviour in
itself cannot be interpreted as a proof for the quantum origin of the
perturbations. However, in more complicated inflationary scenarios it is
possible to find an observable, in principle, feature of the perturbation
spectrum $|f_k|^2$ (namely the absence of zeroes of $|f_k|^2$ as a function
of ${\bf k}$) that might be interpreted as a "quantum signature" though
perturbations themselves are classical nowadays as usual~\cite{david95}.

In addition, the standing wave behavour of $f_k$ yields the appearance
of a component in the ${\Delta T\over T}$ polarization multipoles produced by
the primordial GW background which is periodic in $l$. The period of
oscillations is $T_l=\pi (\eta_0-\eta_{rec})/\eta_{rec}$ in the spatially
flat Universe where $\eta_0$ and $\eta_{rec}$ are the present conformal
time and the recombination time
respectively. The amplitude of this component is, however, on the level
${\Delta T\over T} \sim 10^{-6}$ or less, the corresponding effect in the
$\left({\Delta T\over T}\right)_l$ itself is even smaller.
\par
The background consisting of waves $\propto f_q(k)$ belonging to the quasi-
isotropic mode only, despite its stochasticity, occupies a volume of
measure zero in phase space. Thus, one might expect that its entropy is
low, if any. To determine the volume occupied in phase space, one has to
restore the exponentially small decaying mode, or to consider other physical
processes leading to the generation of both modes. If the quasi-isotropic mode
remains unchanged, the resulting effect may be described as the appearance
of a small stochastic correction to the phases $\xi_k$ and $\theta_k$. There
were a number of proposals about how to calculate the entropy of the
perturbations~\cite{pro,gasperini,oxman}, all of them leading to the same
expression
$\Delta S_k\simeq 2 r_k~(r_k\gg 1)$ for each mode ${\bf k}$ appearing as a
result of coarse graining. It remains unclear, however, if this entropy is the
minimal possible one that has to be ascribed to the perturbations irrespective
of the choice of coarse graining. We hope to return to this question elsewhere.

\vspace{1cm}
\noindent
{\bf Acknowledgements}
\par\noindent
This work was finished when one of the authors (A.S.) visited France under
the agreement between the Landau Institute for Theoretical Physics and
Ecole Normale Sup\'erieure, Paris. A.S. thanks ENS and EP93 CNRS (Tours) for
financial support and Profs E. Brezin, C. Barrabes and N. Deruelle for their
hospitality in ENS, Universit\'e de Tours and Observatoire de Paris-Meudon
respectively. A.S. also acknowledges financial support by the Russian
Foundation for Basic Research, grant 93-02-3631, and by the EC
grant INTAS-93-493.


\begin{thebibliography}{99}
\bibitem{hsg} Hawking~S~W 1982 {\it Phys. Lett. {\bf 115B}} 295; Starobinsky
~A~A 1982 {\it Phys. Lett. {\bf 117B}} 175; Guth~A~H and S-Y~Pi 1982
{\it Phys. Rev. Lett.} {\bf 49} 1110
\bibitem{smoot} Smoot~G~F, Bennett~C~L, Kogut~A, Wright~E~L {\it et al} 1992
{\it Ap. J. Lett.} {\bf 396} L1
\bibitem{staro79} Starobinsky~A~A 1979 {\it Pisma v Zh. Eksp. Teor. Fiz.}~
{\bf 30} 719 [1979 {\it JETP Lett.} {\bf 30} 682]
\bibitem{staro85} Starobinsky~A~A 1985 {\it Pisma Astron. Zh.}~
{\bf 11} 323 [1985 {\it Sov. Astron. Lett.} {\bf 11} 133]
\bibitem{carr} Novikov~I~D, Polnarev~A~G, Starobinsky~A~A and Zeldovich~Ya~B
1979 {\it Astron. Astroph.} {\bf 80} 104; Carr~B~J and Lidsey~J~E 1993
{\it Phys. Rev. D} {\bf 48} 543
\bibitem{grish1} Grishchuk~L~P and Sidorov~Yu~V 1989 {\it Class. Quantum Grav.}
{}~{\bf 6} L161; 1990 {\it Phys. Rev. D} {\bf 42} 3413
\bibitem{grish2} Grishchuk~L~P 1993 {\it Class. Quantum Grav.} {\bf 10} 2449
and references therein.
\bibitem{albrecht} Albrecht~A, Ferreira~P, Joyce~M and Prokopec~T 1994
{\it Phys. Rev.D} {\bf 50} 4807
\bibitem{par69} Parker L 1969 {\it Phys. Rev.} {\bf 183} 1057
\bibitem{zel} Zeldovich~Ya~B and Starobinsky~A~A
1971 {\it Zh.~Eksp.~Teor.~Phyz.} {\bf 61} 2161 [1972 {\it Sov. Phys. - JETP}~
{\bf 34} 1159]
\bibitem{grish3} Grishchuk~L~P 1974 {\it Zh.~Eksp.~Teor.~Phyz.} {\bf 67} 825
[1974 {\it Sov. Phys. - JETP} {\bf 40} 409]
\bibitem{par74} Parker~L and Fulling~S~A 1974 {\it Phys. Rev. D} {\bf 9} 341
\bibitem{hal} Hawking~S~W and Halliwell~J 1985 {\it Phys. Rev. D} {\bf 31} 1777
\bibitem{barut} Barut~A~O and Girardello~L 1971 {\it Comm. Math. Phys.}
{\bf 21} 41
\bibitem{perelomov} Perelomov~A~M 1977 {\it Usp. Fiz. Nauk.} {\bf 123} 23
[1977 {\it Sov. Phys. Usp.} {\bf 20} 703]
\bibitem{schum} Caves~C~M and Schumaker~B~L 1985 {\it Phys. Rev. A} {\bf 31}
3068, 3093; Schumaker~B 1986 {\it Physics Reports} {\bf 135} 317
\bibitem{lifshits} Lifshits~E~M 1946 {\it Zh. Eksp. Teor. Fiz.} {\bf 16} 587
\bibitem{star84} Starobinsky~A~A 1984 {\it Fundamental Interactions}
ed Ponomarev~V~N (Moscow: MGPI Press) p. 55
\bibitem{star86} Starobinsky~A~A 1986 {\it Field Theory, Quantum Gravity
and Strings (Lecture Notes in Physics Vol. 246)} ed de Vega~H~J and
Sanchez~N (Berlin: Springer-Verlag) p. 107
\bibitem{GP85} Guth~A~H and Pi~S~Y 1985 {\it Phys. Rev. D} {\bf 32} 1899
\bibitem{LM93} Laflamme~R and Matacz~A 1993 {\it Int. J. Mod. Phys. D}
{\bf 2} 171
\bibitem{LL} Landau~L and Lifshitz~E~M 1970 {\it Th\'eorie des Champs}
(Moscou:~Editions Mir)
\bibitem{morikawa} Morikawa~M 1990 {\it Phys. Rev. D} {\bf 42} 2929
\bibitem{habib} Halliwell~J 1989 {\it Phys. Rev. D} {\bf 39} 2912; Kiefer~C
1987 {\it Class. Quantum Grav.} {\bf 4} 1369; Habib~S and
Laflamme~R 1990 {\it Phys. Rev. D} {\bf 42} 4056
\bibitem{david95} Polarski~D and Starobinsky~A~A 1995 {\it Phys. Lett.}
{\bf 356B} 196 
\bibitem{pro} Brandenberger~R, Mukhanov~V and Prokopec~T 1992 {\it Phys.
Rev. Lett.} {\bf 69} 3606; Prokopec~T 1993 {\it Class. Quantum Grav.}
{\bf 10} 2295
\bibitem{gasperini} Gasperini~M and Giovannini~M 1993 {\it Phys. Lett.}
{\bf 301B} 334; 1993 {\it Class. Quantum Grav.} {\bf 10} L133
\bibitem{oxman} Kruczenski~M, Oxman ~L~E and Zaldarriaga~M 1994 {\it Class.
Quantum Grav.} {\bf 11} 2377
\end{thebibliography}
\end{document}